\documentclass{sbc2023}%

\usepackage{float}
\usepackage{graphicx}
\usepackage[utf8]{inputenc}
\usepackage[misc,geometry]{ifsym} 
\usepackage{fontawesome}
\usepackage{academicons}
\usepackage{color}
\usepackage{hyperref} 
\usepackage{aas_macros}
\usepackage[bottom]{footmisc}
\usepackage{supertabular}
\usepackage{afterpage}
\usepackage{pifont}
\usepackage{multicol}
\usepackage{multirow}
\usepackage{tikz}
\usepackage{environ}
\usetikzlibrary{positioning, shapes.geometric, arrows.meta}
\usetikzlibrary{arrows}
\setcitestyle{square}

\usepackage{xurl}

\definecolor{orcidlogo}{rgb}{0.37,0.48,0.13}
\definecolor{unilogo}{rgb}{0.16, 0.26, 0.58}
\definecolor{maillogo}{rgb}{0.58, 0.16, 0.26}
\definecolor{darkblue}{rgb}{0.0,0.0,0.0}
\hypersetup{colorlinks,breaklinks,
            linkcolor=darkblue,urlcolor=darkblue,
            anchorcolor=darkblue,citecolor=darkblue}

\jtitle{}
\jyear{2024}

\title[The trade-offs between Monolithic vs. Distributed Architectures]{The trade-offs between Monolithic vs. Distributed Architectures}

\author[Felisberto. 2024]{
    \affil{\textbf{Matheus Felisberto}~[~\textbf{FIA Business School}~|~\href{mailto:matheus.felisberto@gympass.com}{\textbf{\textit{matheus.felisberto@gympass.com}}}~]}
}

\begin{document}

\begin{frontmatter}
\maketitle

\begin{abstract}
\textbf{Abstract}
\noindent Software architects frequently engage in trade-off analysis, often confronting sub-optimal solutions due to unforeseen or overlooked disadvantages. Such outcomes can detrimentally affect a company's business operations and resource allocation. This article conducts a critical review of architectural styles, particularly focusing on the strengths and weaknesses of both monolithic and distributed architectures, and their relationship to architectural characteristics. It also explores the role of cloud computing in transitioning from monolithic to distributed-based applications. Utilizing a broad range of sources, including papers and books from both industry and academia, this research provides an overview from theoretical foundations to practical applications. A notable trend observed is a shift back from distributed to monolithic architectures, possibly due to factors such as cost, complexity, and performance.
\end{abstract}


\begin{license}
Published under the Creative Commons Attribution
\end{license}

\end{frontmatter}

\section{Introduction}
\label{sec:intro}

As technology evolves in a fast-paced environment, cloud computing has emerged significantly reshaping how software is designed, deployed, and utilized. This shift towards a cloud-native era, favoring agility, scalability, and elasticity, represents a move away from traditional monolithic, on-premises solutions. In conjunction, new architectural styles like serverless computing, have become a viable and popular alternative for many businesses.

However, an often-overlooked aspect is the trade-offs involved. The complexities introduced by these architectural choices are usually challenging to modify, as per the definition of software architecture by \citep{peaa}. The trade-offs may range from underutilizing a system, such as a database management systems, to dealing with the intricate complexities of a distributed transaction. 

This article aims to provide a systematic review of architectural styles exploring both benefits and inherent challenges between monolhitc and distributed-based architectures in what might be described as the no-free-lunch theorem scenario in software design.

\section{Methodology}
This article employs a methodology that blends snowballing research with a grey literature review, focusing on architectural styles and characteristics between distributed architectures and monoliths. This study began with a search for foundational papers and reports in these fields, using academic databases and industry publications. Through the snowballing process, the base literature collection was progressively expanded. To ensure a comprehensive coverage, I traced backward and forward the references, capturing both seminal and recent works. Supplementing these academic findings, white papers, technical reports, blog posts and industry case studies were also reviewed to provide a practical industry insights. 

The gathered materials were deeply analyzed to identify common themes, trends and divergences in the field. This phase involved extracting and synthesizing key insights related to the research themes, leading to the findings presented in this article.
In conjunction snowballing research with grey literature review, enabled a thorough and multifaceted exploration of the topics, capturing both theoretical and practical insights.

\section{Software Architecture}
Software Architecture plays a crucial role in the industry, however, it still lacks a descriptive and universally convincing definition. Over the years, many experts have attempted to delineate its boundaries, often in the presence of exceptions and caveats. It can be thought of as the composition of a system's components and the relationships between them. The primary focus is on the public interface; the architectural significance lies in these externally facing elements rather than in the private details. However, even though they are not the the main concern, can still play a role in shaping the overall structure and functionality of the system, underscores \cite{sap}.

Building software on a solid foundation, which might include robust architectural patterns and coding standards, is essential, especially considering the key role it plays in a business’s success. Failing to address a wide range of potential scenarios could significantly mining a company’s growth. While many modern frameworks offer solutions to common problems, supporting in these architectural decisions, it’s important to recognize that each system has its own unique requirements and specificities that must be carefully considered, indicates \cite{maag}.

\subsection{Architectural Characteristics}
\label{sec:ac}
Every software inherently depends on factors beyond its immediate domain, which architects must consider during the design process. These factors, often  referred to as Quality Attributes or non-functional requirements, have been studied at least since 1970s by the software community, portrays \cite{sap}, and are essential for ensuring the software meets both business and user needs. Extending from low-level code aspects, such as modularity, to operational concerns, including scalability, as detailed in \cite{sapf} and \cite{fosa}. It is important to recognize that every software characteristics list, is an incomplete, ambiguous, and overlapping list, regardless of the very own \cite{iso}. For instance, \cite{fosa} exemplifies that interoperability and compatibility might be equivalent depending on the system. \cite{sap} also point out a particular case of denial-of-service attack. Would it be an aspect of availability, performance, security, or usability? In some extent, we can consider them all.
\leavevmode
One of the main reasons behind the imprecise definitions and the persisting ambiguity, is the fast-pacing evolution in the industry itself. With many different use cases, companies develop their own terminologies, complicating efforts to establish comprehensive taxonomies for all categories. However, it is recommended to adopt the Domain Driven Design, specifically the ubiquitous-language technique, to minimize misunderstandings. As outlined by \cite{ddd}, this approach emphasizes consistency in both language and terminology across the whole domain, thereby fostering a better comprehension and clearer communication. 

\begin{quote}
    \textit{Domain experts should object to terms or structures that are awkward or inadequate to convey domain understanding; developers should watch for ambiguity or inconsistency that will trip up design.} \cite{ddd}
\end{quote}

When it comes to architectural characteristics, it is essential to understand that not every application will require or support every characteristic maximization. It is a decision-making process selecting the aspects that are relevant to solve a problem at hand, while taking into consideration the impact among them. Take, for example, a system where security is crucial. By prioritizing this characteristic, it might cause implications on performance, such as the need for encryption at-rest and at-transit, and secure storage strategies for sensitive data. Attempting to excel in every aspect can lead to the development of a generic architecture, which, in its naively attempt to address a wide range of business problems, may lack focus and efficiency. 

\cite{fosa} advocate for an iterative design process. This approach suggests that implementing smaller, iterative changes is more effective than trying to preemptively address every possible scenario, which can lead to over complexity and unanticipated challenges.

\begin{quote}
    \textit{Never shoot for the best architecture, but rather the least worst architecture.} \cite{fosa}
\end{quote}

\subsubsection{Availability}
\label{sec:availability}
Availability refers to the aspect of software being ready and operational when needed. Given the complexity of systems, and the wide range of potential, and unavoidable issues, maintaining high availability is a critical challenge. This involves not only identifying failures cause by faults but also implementing repair mechanisms, ideally autonomous, that is, without the need for human intervention. Among the difficulties in designing highly available, fault-tolerant systems is accurately predicting potential failure modes and then devising effective strategies to mitigate them. As \cite{sap} emphasize, this proactive approach to failure detection and mitigation is key to achieving and maintaining system availability.

To quantify a system's availability, Service Level Agreements (SLAs) are typically established. These agreements specify the guaranteed performance standards and outline the penalties for failing to meet these criteria. To illustrate it, the Table \ref{tab:aws-sla} shows the SLAs tiers by the public cloud provider \cite{awsav}.

The challenge lies in not designing with the false and tempting notion of avoiding problems entirely, but rather in considering failure as first-class citizens. By adopting this mindset, it becomes possible to identify the types of failures a system is prone to and to implement appropriate techniques to handle them, as highlighted \cite{sap}. 

\begin{table}
    \centering
    \begin{tabular}{cc}
    \hline \textbf{SLA (\% Uptime)} & \textbf{Unavailability per Year} \\ \hline 
        99\%                        & 3 days 15 hours \\
        99.9\%                      & 8 hours 45 minutes \\
        99.95\%                     & 4 hours 22 minutes \\
        99.99\%                     & 52 minutes \\
        99.999\%                    & 5 minutes \\ \hline
    \end{tabular}
    \caption{AWS availability tiers}
    \label{tab:aws-sla}
\end{table}

\subsubsection{Interoperability}
The interoperability refers to the ability of a different systems to exchange and effectively interpret information. This concept involves two distinct layers: syntactic interoperability, which is concerned with the structure and format of the data exchange, and semantic interoperability, which relates to the meaning and interpretation of the data. These layers are comprehensively described in \cite{sap}. \cite{cpi} further define interoperability as the capacity of a collection of communicating entities to share specified information, and operate on that information according to an agreed operational semantics.
An integral part of achieving interoperability is the process by which systems locate each other and manage their interfaces to facilitate the sharing of information.

\subsubsection{Modifiability}
Software is a continuous work in progress, whether it is already in production or not. Continuous development, ranging from refactoring and adding new features to updating dependencies, is a norm. This could involve upgrades to the runtime, framework, or even the hardware itself. According to \cite{sap}, there are two main tactics to improve a system's modifiability: enhancing cohesion and reducing coupling. Improving the former and minimizing the latter can significantly boost this characteristic.
A notable contribution to this field is by \cite{tstms}, who presented the Technische Hogeschool Eindhoven (THE) operating system. This work was an early example of a well-structure operating system, and introduced the hierarchical model, a novel approach at the time. This approach promotes a better modifiability, as each layer is designed with distinct responsabilities, abstraction levels, control flows, and modular structures.
In the same vein, \cite{cudsm} heavily influenced software design towards modularization, introducing the concept of "information hiding". According to Parnas, modules should have a well-defined interface and hide its working from the rest of the system. This philosophy enables individual modules modifications without affecting others, provided the interfaces remain consistent. Parnas also emphasized that modules are likely to suffer changes throughout their lifecycle, highlighting the importance of reducing inter-module dependencies and hiding information.  

\subsubsection{Performance}
According to the \cite{iso}, performance can be understood through three primary aspects: time behavior, resource utilization, and capacity. These elements collectively define how efficiently a software performs a specific computational task within predetermined time constraints and throughput requirements, while also optimizing resource consumption. The field of performance is broad and has been the subject of extensive research. It necessitates a deep understanding of both hardware and software interactions. As \cite{sps} describes, when the performance of one component is pushed to its limits, the other often plays a critical compensating role.

Performance its particularly interesting as an architectural characteristic because it is directly perceived by the user, as much as subsection \ref{sec:availability}. Consider, for instance, a trading system suffering by latency or an online game where commands are delayed. Such frustrations can drive users away. Performance can be measured in terms of throughput or latency, depending on the application's requirements. In a web server handling thousands of requests, achieving high throughput is desirable. Conversely, online gaming demands low latency. Throughput refers to the amount of data a system can process in a given period, whereas latency measures the time it takes for a request to travel from sender to receiver and back. Improving performance might involve reducing demand, such as refusing to serve requests in an overwhelmed web server, or managing resources more effectively through scheduling, replication, or increasing the resource pool, as explained by \cite{sap}.

\subsubsection{Security}
Is well known that data has become as valuable as oil in the modern era. In the context of software, security is a critical characteristic that measures a system's ability to protect its data, particularly from unauthorized access. According to \cite{sap}, there are three main aspects of security: confidentiality, integrity, and availability. Confidentiality ensures that data is accessible only to authorized users. For example, only you should have access to your mobile banking application. Integrity involves safeguarding data from unauthorized modifications ensuring that any alteration is done only through legitimate means. Finally, availability refers to the system being accessible and usable when legitimate users need it.

In the recent years, regulations like the General Data Protection Regulation (GDPR) have been established to address growing privacy concerns. A key aspect of these regulations is determining which information can be shared and with whom. This decision-making process is crucial, not just regarding what information is shared, but also the nature of the information being shared. Consider this scenario: you forget your password and initiate a process to recover your mobile account. You enter your email and set a new password. In the event of a data breach, you can change your email or password, albeit with some inconvenience. But what if the leaked information included your biometric data?

Security is about more than just encrypting data, whether at-rest or in-transit. It also involves meticulously tracking each piece of data, understanding why it’s needed, and ensuring traceability, as demanded by regulatory bodies. This comprehensive approach to security helps protect sensitive information, especially in cases where it cannot be easily changed or replaced, like biometric data.

\subsubsection{Testability}
To ensure that software is functioning correctly and ideally free of known faults or bugs, it's important to understand what constitutes a software bug. \cite{st} outlines at least five scenarios that can be considered as software bugs:

\begin{enumerate}
    \item \textit{The software fails to do something that the product specification states it should do.}
    \item \textit{The software does something that the product specification explicitly says it should not do.}
    \item \textit{The software performs an action that the product specification does not cover.}
    \item \textit{The software omits an action that is not mentioned in the product specification but is implied or necessary.}
    \item \textit{The software is user-unfriendly, slow, or appears flawed to the software tester or the end user}.
\end{enumerate}

Testability, the ease with which software can be tested, is crucial in this context. It refers to how readily a software system, either in isolation or in combination, can be tested for bugs. This is often facilitated through test harness tools that control inputs, outputs, and possibly the state of the software, ensuring it aligns with the provided specifications. High testability is linked to increased reliability, making it a valuable characteristic to consider in software architecture.

\subsubsection{Usability}
Usability is a measure of how easily users can interact with a system. Its significance has increased over the years, becoming a critical factor in users’ decision-making when choosing between services. \cite{sap} describes several dimensions of usability. One key aspect is the ease of learning system features, often referred to as onboarding. For example, how quickly can a new user learn to use the system? Implementing short tutorials is a useful tactic in this regard.
Efficient use of the system and minimizing the impact of errors are also crucial. The former could involve allowing users to pause and later resume a long-running task. The latter ranges from displaying errors in a user-friendly manner to providing recovery options, like retrying or suggesting alternative steps to achieve the same objective.
Adapting the system to meet user needs is another way to enhance usability. A familiar example is a browser auto-filling credit card details based on previous usage. Lastly, boosting users confidence and satisfaction also plays a role in enhancing usability. For instance, providing visual feedback during a video upload, like showing a progress percentage, can significantly improve the user experience.

\subsection{Architectural Styles}
Architectural styles represent the macro structure of a system and are categorized based on a set of characteristics, each offering distinct advantages and disadvantages. One of the key responsibilities of an architect is to select the appropriate architectural style by evaluating it against the business requirements. The following sections systematically describe both monolithic and distributed-based architectural styles in a non-exhaustive manner.

\subsubsection{Layered Architecture}
\label{sec:la}
The layered architecture is one of the most straightforward and commonly used architectural styles, appreciated for its simplicity, popularity, and cost-effective implementation, as described by \cite{fosa}. This architectural style naturally aligns with the needs and structures of many traditional business applications.

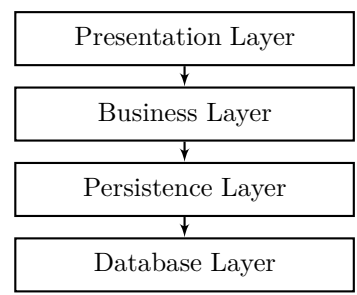
\begin{figure}[h]
\centering
\begin{tikzpicture}[scale=0.7, auto, node distance=1cm, thick, align=center,
    box/.style={rectangle, draw, text width=12em, minimum height=2em},
    line/.style={draw, -latex'}]

    \node[box] (presentation) {Presentation Layer};
    \node[box, below of=presentation] (business) {Business Layer};
    \node[box, below of=business] (persistence) {Persistence Layer};
    \node[box, below of=persistence] (database) {Database Layer};

    \path[line] (presentation) -- (business);
    \path[line] (business) -- (persistence);
    \path[line] (persistence) -- (database);
\end{tikzpicture}
\vspace{10pt}
\caption{Layered-architecture style topology}
\label{fig:las}
\end{figure}

The topology of a layered architecture style, as illustrated in Figure \ref{fig:las}, is characterized by logical horizontal layers, each assuming a specific role within the application. While there is no strict limitation on the number of layers, four standard ones are commonly observed: presentation, business, persistence, and database. The actual number and nature of these layers can vary depending on the size and complexity of the application.

Each layer in a layered architecture has a distinct role and responsibility. For example, the business layer handles the application's business logic, while the presentation layer manages browser communication and interface logic. This structure ensures that each layer abstracts its specific function to collectively fulfill a business request. For instance, the presentation layer doesn't concern itself with customer details, which are the purview of the business layer, nor does the business layer need to manage HTML markup. This separation of concerns facilitates the construction of layered architectures with clearly defined responsibilities for each layer. However, as \cite{fosa} point out, one drawback of this approach is the challenge in applying domain-driven design. In layered architectures, the business domain tends to be dispersed across all layers, categorized by technical role rather than being grouped into components like Customer.

\paragraph{Suggested applications}
The layered architecture is a great choice in the early stages of a project, particularly when there is significant uncertainty about the product itself but development needs to begin. As \cite{sap2} details, it is also well-suited for projects with time or budget constraints. Being a monolithic style, it avoids the complexities associated with distributed systems, such as the need to define contracts and manage remote procedure calls. These aspects of distributed systems will be discussed in more detail in Section \ref{sec:fds}.

\paragraph{Discouraged applications}
One important consideration with layered architecture is its scalability challenge. As a monolithic architectural style, scaling up for improved performance, elasticity, or overall scalability typically involves scaling the entire application. This approach may not always be efficient. Furthermore, layered architectures may lack sufficient fault tolerance. Since all components are deployed together, a fault in one area can potentially bring down the entire system, as \cite{sap2} notes.

\subsubsection{Microkernel architecture}
Inspired by operating systems, the microkernel architecture style is known for its flexibility and extensibility through the use of plugins, often referred to as plug-in architecture. Although it was developed several years ago, it remains widely used today. Its extensibility lies in maintaining core functionality while allowing for the addition of varied features through plugins. This architecture comprises two fundamental components: the core and plugin modules. The core provides essential functionalities, whereas plugin modules add additional, usually standalone, independent components. A common method for the core system to recognize available plugins is through a plugin registry, which contains all necessary information about them. While these plugins have traditionally been developed as libraries, they can also be implemented through remote services such as REST, according to \cite{sap2}.

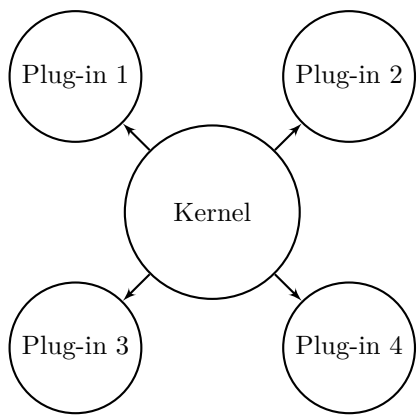
\begin{figure}[h]
\centering
\begin{tikzpicture}[scale=0.7, auto, node distance=0.5cm, thick, align=center,
    module/.style={circle, draw, minimum size=1cm},
    microkernel/.style={circle, draw, minimum size=2.3cm},
    line/.style={draw, -latex'}]

    \node[microkernel] (microkernel) {Kernel};

    \node[module, above left=of microkernel] (plugin1) {Plug-in 1};
    \node[module, above right=of microkernel] (plugin2) {Plug-in 2};
    \node[module, below left=of microkernel] (plugin3) {Plug-in 3};
    \node[module, below right=of microkernel] (plugin4) {Plug-in 4};

    \path[line] (microkernel) -- (plugin1);
    \path[line] (microkernel) -- (plugin2);
    \path[line] (microkernel) -- (plugin3);
    \path[line] (microkernel) -- (plugin4);

\end{tikzpicture}
\vspace{10pt}
\caption{Microkernel architecture style topology}
\label{fig:mkas}
\end{figure}

\paragraph{Suggested applications}
This architectural style is an excellent choice for products that require customization or will be expanded with more features over time, providing an easy starting point for such projects. The microkernel architecture also adapts well to varying configurations or deployment requirements. For instance, a microkernel application can be deployed in an on-premise facility with a custom set of plugins tailored to specific needs, while maintaining core functionality that remains unchanged and completely agnostic. \cite{sap2} emphasizes that the microkernel, as well as discussed in subsection \ref{sec:la}, offers a cost-effective and relatively simple setup, making it a practical option for projects with limited time or budget.

\paragraph{Discouraged applications}
The downsides of this architectural style lie in its heavy reliance on the core module, which can act as a bottleneck. This dependency complicates elasticity, reduces fault tolerance, and is less suited for highly scalable solutions. Additionally, the architecture discourages frequent modifications to the core module, preferring instead to use plugins to add features and extend functionality. When frequent core updates are necessary, this approach becomes inadequate. As \cite{sap2} suggests, if adhering to this rationale is not feasible, this architectural style may not be the best solution to the problem at hand.

\subsubsection{Event-Driven architecture}
Event-driven architecture is a distributed, asynchronous architectural style that has become increasingly popular, especially following the democratization of the cloud by public vendors. This architecture excels in managing complex workflows and fostering reactive, responsive systems. It has also boosted the creation of tools, frameworks, and cloud-based solutions tailored to enhance accessibility and ease of setup.

In an event-driven system, decoupled event processing components receive and process events asynchronously. The process is initiated by an event, which kicks off an asynchronous workflow. For example, consider placing an order on an e-commerce platform that uses event-driven architecture. The moment you place the order, an event is created containing all necessary details. This event triggers the asynchronous workflow. Subsequently, a component, such as the payment processor, receives this event, processes it, and may generate another event to inform the system of a state change. Once the payment is confirmed, another event is published, prompting the shipment component to act. All these interactions involve physical messaging artifacts. Typically, initiating events utilize point-to-point channels through messaging services or queues, while processing events often employ publish-subscribe channels via topics or notification services, as described by \cite{sap2}. Event-driven architecture can function as a standalone architectural style or be combined with others, such as microservices.

\begin{figure}[h]
\centering
\begin{tikzpicture}[scale=0.9, auto, node distance=0.5cm and 1.0cm, thick, align=center,
    event/.style={rectangle, draw, text width=5.1em, text centered, minimum height=2em},
    processor/.style={rectangle, draw, text width=5.1em, text centered, minimum height=2em},
    channel/.style={cylinder, shape border rotate=0, draw, text width=4em, minimum height=3em, shape aspect=0.3},
    arrow/.style={thick,->,>=stealth}]

    \node[event] (event) {Event};
    \node[channel, right=of event] (channel) {Channel};
    \node[processor, above right=of channel] (processor1) {Processor 1};
    \node[processor, right=of channel] (processor2) {Processor 2};
    \node[processor, below right=of channel] (processor3) {Processor 3};

    \draw[arrow] (event) -- (channel);
    \draw[arrow] (channel) -- (processor1) node[midway, above] {};
    \draw[arrow] (channel) -- (processor2) node[midway, right] {};
    \draw[arrow] (channel) -- (processor3) node[midway, below] {};

\end{tikzpicture}
\vspace{10pt}
\caption{Event-driven architecture style topology}
\label{fig:eda}
\end{figure}
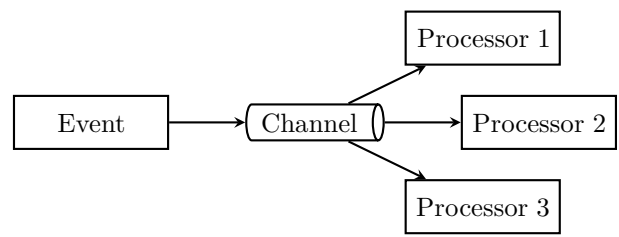

\paragraph{Suggested applications}
Event-driven architecture is undoubtedly an optimal choice for achieving high performance, scalability, and fault tolerance. It also brings additional benefits as it shifts the paradigm from traditional decision-tree logic to a model that reacts to events. This approach facilitates the modeling of highly complex systems by isolating each component within its context, focusing on its specific responsibilities, and processing incoming events. Moreover, it can aid organizational scaling by allowing teams to work independently on their components, each defined by clear contracts. However, the discussion of team typologies is out of scope of this article.

\paragraph{Discouraged applications}
By its nature, being a distributed asynchronous architecture, the event-driven architectural style is not recommended for request-based systems that require high data consistency, as revealed by \cite{sap2}. Due to its asynchronous nature and the eventual consistency of processing, meaning there is no guarantee of when an event will be processed, event-driven architecture may not suit scenarios demanding immediate data accuracy, such as CRUD operations.

Additionally, managing the order in which events are consumed and handling errors present significant challenges. Sequencing issues can complicate the architecture, especially if events need to be combined to produce further actions, or the potentially deadlock situations where, for instance, event A waits for event B, event B waits for event C, and event C waits for event A. Error handling adds another layer of complexity. For example, consider a scenario where a payment event is processed successfully, but a subsequent inventory event fails because the item is out of stock. In such cases, stakeholders must clearly define responsibilities and corrective actions for each potential issue, which can lead to corner cases that compromise the system's reliability in terms of data consistency.

\subsubsection{Microservice architecture}
Microservice architecture is arguably the most popular and widely used architectural style today. The term 'microservice' was coined in 2014 by Martin Fowler and James Lewis. Although they were the first to publish an article detailing its characteristics, they emphasized that this architectural style was not a novelty but rather rooted in UNIX philosophy. Additionally, microservices incorporate Domain-Driven Design concepts, such as bounded contexts, which organize entities and behaviors within well-defined limits in code and database schemas. Microservices operate by integrating multiple smaller, independently deployed services that communicate with each other. This approach is defined by Fowler and Lewis as follows:

\begin{quote}
\textit{In short, the microservice architectural style is an approach to developing a single application as a suite of small services, each running in its own process and communicating with lightweight mechanisms, often an HTTP resource API. These services are built around business capabilities and independently deployable by fully automated deployment machinery.} \cite{ms}
\end{quote}

The primary goal of microservice architecture is to achieve decoupling, with physical modeling that represents bounded contexts. To accomplish this, each service is designed to be self-contained, possessing all necessary components for operation, including its own database systems. This approach eliminates a single source of truth, allowing different services to utilize diverse technologies tailored to their specific needs. For example, in contrast to a monolithic architecture like a subsection \ref{sec:la}, where performing a full-text search using a traditional RDBMS might be unsatisfactory at scale, a microservice architecture could leverage a specialized Elasticsearch or Solr cluster to handle such tasks more effectively.

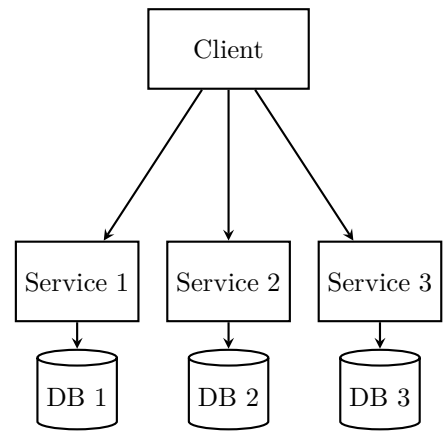
\begin{figure}[h]
\centering
\begin{tikzpicture}[scale=0.5, thick, align=center,
    service/.style={rectangle, draw, minimum width=4em, text centered, minimum height=3em},
    client/.style={rectangle, draw, minimum width=6em, text centered, minimum height=3em},
    database/.style={cylinder, shape border rotate=90, draw, minimum width=2.5em, minimum height=3em, shape aspect=0.2},
    arrow/.style={thick,->,>=stealth},
    node distance=1em]

    \node[client] (client) {Client};

    \node[service, below=2cm of client] (service2) {Service 2};
    \node[service, left=of service2] (service1) {Service 1};
    \node[service, right=of service2] (service3) {Service 3};

    \node[database, below=of service1] (db1) {DB 1};
    \node[database, below=of service2] (db2) {DB 2};
    \node[database, below=of service3] (db3) {DB 3};

    \draw[arrow] (client) -- (service1);
    \draw[arrow] (client) -- (service2);
    \draw[arrow] (client) -- (service3);
    \draw[arrow] (service1) -- (db1);
    \draw[arrow] (service2) -- (db2);
    \draw[arrow] (service3) -- (db3);

\end{tikzpicture}
\vspace{10pt}
\caption{Microservice architecture style topology}
\label{fig:ms}
\end{figure}

As a distributed architecture, each service in a microservice setup must run independently in its own process. A significant factor in the widespread adoption of this architectural style has been the rise of cloud computing, which has made managing these physical constraints easier through virtualization and containers. According to \cite{fosa}, decoupling services to the extent that each domain maintains its own infrastructure was once impractical. However, with the availability of free open-source operating systems and automated infrastructure management, it is now feasible at both the domain and operational levels.

Network calls present a bottleneck in this architecture, as they are typically slower than straightforward method calls due to the process of transferring data across a network, which often includes a layer of security. Additionally, communication between microservices requires that each service knows how to initiate contact with others without relying on a centralized orchestrator, to prevent coupling. \cite{fosa} characterizes this communication as protocol-aware, meaning the caller must know the communication protocol in advance, and heterogeneous, allowing services to be developed using different technologies, thus ensuring interoperability.

\paragraph{Suggested applications}
When elasticity, high fault tolerance, and scalability are priorities, the microservice architecture style is an excellent choice. Extensibility is another significant advantage; adding functionality typically involves spinning up new services within the already standardized infrastructure. This flexibility is crucial, allowing rapid adaptations and changes aligned with business needs. The architecture also enhances maintainability due to each service's smaller scope and the clear boundaries guaranteed by the concept of bounded contexts. Testing is generally more straightforward, as long as dependencies on other services are minimized. When dependencies are unavoidable, a robust mocking strategy can effectively facilitate testing.

\paragraph{Discouraged applications}
The primary goal of microservices is decoupling. However, this architecture may not be suitable in scenarios where, despite physical decoupling, there remains substantial logical coupling. This is particularly problematic in systems with complex workflows and excessive inter-service communication. Often, the need for frequent communication between services is driven by data dependencies, which are a critical factor in deciding whether to adopt this architectural style. The more tightly coupled the data, the less effective the microservice architecture becomes.

Microservices represent one of the most complex architectures in use today. They can also be costly due to the duplication of efforts, including data replication across multiple databases. As the number of services increases, the infrastructure required for each service also grows exponentially, as delineated by \cite{fosa}.

Despite what may seem counter intuitive, microservices are not inherently high-performance or highly-responsive systems. Their performance is often hampered by three types of latency: network latency, security checks, and data access delays. Each of these factors needs to be fine tuned to the specific needs of each service, even at the database level.

\section{Architectural decisions}
One of the key responsibilities of a software architect is to analyze a range of factors that differ from business to business and technology to technology. Unfortunately, there is no one-size-fits-all solution. Choosing an architectural style requires careful consideration of trade-offs, ensuring that the disadvantages are not overlooked in favor of the advantages. Many factors contribute to this necessity. One significant factor is the hype surrounding new frameworks, language features, or even new programming languages. It is natural for developers and technologists to be excited by the purported innovations these new tools bring. However, this enthusiasm can sometimes lead to solutions that are overly complicated, costly, and complex. \cite{sathp} describe this phenomenon as evangelism, where the focus on the positive aspects leads to neglecting potential downsides.

In order to mitigate risks while ensuring that the architecture evolves in line with business requirements, it is crucial to analyze and prioritize what is most important at any given time, or to project future needs as discussed in section \ref{sec:ac}. Unfortunately, distributed architectures, particularly microservices, are often default choices for many applications. This trend is driven by misconceptions and an underestimation of the trade-offs involved. The widespread popularity of this architectural style is not coincidental, and it necessitates a careful consideration of its implications.

Recently, even big tech companies, equipped with their own data centers and substantial intellectual capital, have begun shifting toward monolithic architectures. This move aims to reduce costs, complexity, and scalability issues. \cite{tmdca} propose a programming methodology that reportedly reduces latency by 15 times and costs by 9 times compared to the current standard. The paper also critiques some commonly touted benefits of microservices, pointing out how they can actually lead to problems in performance, correctness, and development agility. For example, performance can suffer when the granularity of services is poor, leading to excessive inter-service communication. Correctness becomes challenging when considering all possible interactions between services, including variations in service versions and handling failures. Development agility is compromised when changes across services cannot be made atomically, requiring extensive coordination effort. To address these issues, the proposed methodology advocates for writing monolithic applications that are modularized as components, assigning these components to physical processes, and deploying them atomically.

Another compelling example of cost reduction achieved by shifting from a distributed microservices-based architecture to a monolith is the Amazon Prime Video case. This case study describes how the system was initially orchestrated by a component that not only created a bottleneck but also incurred the highest costs within the solution. After transitioning to a monolithic architecture, the infrastructure costs were reduced by an astonishing 90\%, and it is also reported to have enhanced the system's scalability, as described by \cite{spvrc}.

\subsection{Fallacies of distributed systems}
\label{sec:fds}
A common misconception about transitioning from a monolithic to a distributed architecture is encapsulated in the 8 fallacies of distributed systems. For instance, a local method call used to process a business rule is not equivalent to inter-service communication that relies on the network. Numerous precautions are necessary in such environments, including robust error handling. While TCP is a reliable protocol that will resend any lost packets, higher levels in the stack, such as those involving REST APIs, still require individual request management. Additionally, latency increases are a concern. Even though the TCP slow start can be mitigated by establishing long-lived connections, the overhead from serialization and deserialization, along with additional security checks, can significantly extend the final processing time.

\begin{enumerate}
    \item \textbf{The network is reliable:} Assumes that network connections are always stable and reliable.
    \item \textbf{Latency is zero:} Ignores the time delay in communication over a network.
    \item \textbf{Bandwidth is infinite:} Ignores the limitations in network data transfer capacity.
    \item \textbf{The network is secure:} Assumes that the network is naturally secure from attacks and vulnerabilities.
    \item \textbf{Topology does not change:} Assumes that the way network systems are connected remains constant.
    \item \textbf{There is one administrator:} Ignores the complexity of managing distributed systems with multiple administrative domains.
    \item \textbf{Transport cost is zero:} Ignores the resources and time required to move data across the network.
    \item \textbf{The network is homogeneous:} Assumes that the network’s hardware, software, and protocols are consistent and compatible.
\end{enumerate}

\subsection{Data management}
\label{sec:data-management}
Data management is undoubtedly one of the most complex and critical aspects to consider when comparing monolithic and distributed architectures. In a monolithic architecture, data management is generally simpler because the same system both writes to and reads from a single database. This setup naturally supports transactional capabilities, ensuring that ACID (Atomicity, Consistency, Isolation, Durability) properties are maintained. There is less concern about data integrity, consistency, and security since authentication and authorization are more straightforward in a unified environment. Querying data is also relatively simple, although it may require complex join clauses that necessitate the creation of views. Nevertheless, this setup is typically sufficient for the needs of most small to mid-sized businesses.

Conversely, the more distributed an architecture becomes, the more challenging it is to maintain these properties. In such environments, ensuring data integrity, consistency, and security across multiple services and databases can be daunting. It is usually inadvisable — except in exceptional cases, to attempt to manage these aspects manually due to the increased complexity and risk involved.

\subsubsection{Transactional complexity}

Transactions are a fundamental aspect of enterprise applications, with ACID properties helping to shield developers from many concerns related to data integrity. In monolithic architectures, these properties are typically leveraged without additional complications, as the architecture does not introduce inherent challenges in managing data consistency. However, distributed architectures, particularly those utilizing a database-per-service pattern, can encounter difficulties when transactions need to update data distributed across multiple services.

To address this, the Saga Pattern is employed, which orchestrates a sequence of local, message-driven transactions to ensure data consistency across services, as detailed in \cite{mp}. Yet, implementing the Saga Pattern introduces its own challenges, such as no isolation which necessitates countermeasures to manage concurrency issues effectively. Sagas can be managed through two approaches: choreography, where each local transaction publishes events that trigger subsequent transactions in other services, and orchestration, where a central coordinator directs each participant to execute their transaction. In cases of failure, the saga must initiate compensating transactions in reverse order to undo the effects, as illustrated in \cite{mp}.

\begin{figure*}[tb]
\begin{center}
\begin{tikzpicture}[scale=0.95, transform shape, every node/.style={scale=0.95},
    transaction/.style={rectangle, draw, text width=0.2\textwidth, text centered, minimum height=3em},
    compensating/.style={rectangle, draw, text width=0.2\textwidth, text centered, minimum height=3em},
    arrow/.style={thick,->,>=stealth},
    reverearrow/.style={thick,<-,>=stealth},
    node distance=3cm]

    \node[transaction] (t1) {Transaction 1};
    \node[transaction, right=of t1] (t2) {Transaction 2};
    \node[transaction, right=of t2] (t3) {Transaction 3};
    \node[compensating, below=1cm of t2] (c2) {Compensate T2};
    \node[compensating, below=1cm of t1] (c1) {Compensate T1};

    \draw[arrow] (t1) -- (t2) node[midway, above] {Event 1};
    \draw[arrow] (t2) -- (t3) node[midway, above] {Event 2};
    \draw[arrow] (t2) -- (c2) node[midway, right] {Failure};
    \draw[arrow] (c2) -- (c1) node[midway, below] {Compensate};

\end{tikzpicture}
\caption{Saga pattern topology}
\label{fig:sp}
\end{center}
\end{figure*}
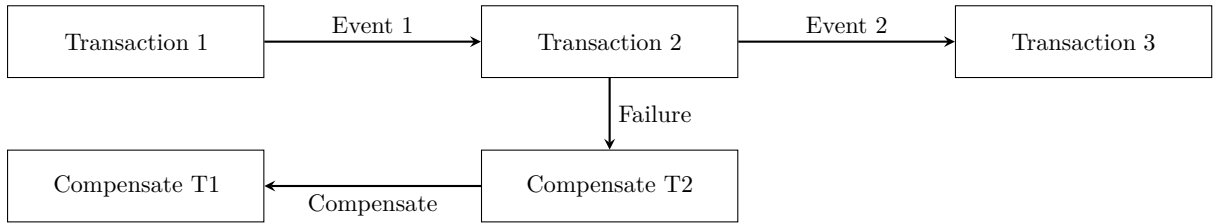

\subsubsection{Eventual Consistency}
To enhance architectural characteristics such as availability and performance, adopting eventual consistency is a strategic decision. \cite{sathp} identifies three main patterns that facilitate this approach: background synchronization, orchestrated request-based pattern, and event-based pattern.

\paragraph{Background Synchronization Pattern} This pattern employs an external system to periodically check and update all data sources to ensure they remain synchronized. The level of eventual consistency achieved depends on the frequency of the updates, which could be scheduled as nightly batch jobs or on an hourly basis. Although this approach effectively keeps data in sync, it also tends to couple related services together. This coupling occurs because the pattern requires knowledge of all data schemas, which can lead to the violation of bounded contexts, duplication of business logic, and a complex implementation.

\paragraph{Orchestrated Request-Based Pattern} Unlike the Background Synchronization Pattern, the goal of this pattern is to process the entire distributed transaction within the scope of the business request itself. It can be implemented either directly within a specific service or through a dedicated orchestrator service. Implementing it within a service tends to overload that service's responsibilities, as it must manage business rules and handle errors in addition to its regular duties. The key issue with this pattern as an eventual consistency approach is its error handling, if an error occurs during the transaction or during a compensating transaction, there will be no external system available to recover the state.

\paragraph{Event-Based Pattern} This pattern utilizes asynchronous publish/subscribe or messaging systems to post events or commands. In this approach, all services are decoupled, and the eventual consistency time factor is minimized due to the asynchronous and parallel nature of the messaging systems, as indicated in \cite{sathp}. The main challenges, similar to other patterns, revolve around error handling. While the messaging system does not require the consumer to be available at the time the message is published — ensuring they receive it when they are available, failures during the processing of these events can still lead to consistency issues.

\subsubsection{Querying data}
Monolithic architectures typically rely on a single database, which generally simplifies the task of querying data. However, in distributed applications, querying becomes significantly more challenging due to the complexities involved in managing data across multiple services. \cite{sathp} outlines four patterns specifically designed to address these challenges.

\paragraph{Interservice Communication Pattern} This pattern is commonly utilized to access data from another service, typically through REST APIs or remote procedure calls such as gRPC. However, it introduces several downsides, including those mentioned in subsection \ref{sec:fds}, and negatively impacts the availability due to service coupling. Additionally, this pattern often requires complex error handling strategies. For example, if a request from service A to service B fails due to a transient error, it directly affects the user experience at that moment.

To mitigate such issues, the Retry Pattern, as defined in \cite{rp}, can be implemented. This pattern automatically retries recoverable failures a specified number of times with exponential backoff, enhancing usability and availability. However, it also risks exacerbating issues in the event of non-recoverable errors, such as a database outage, potentially creating a cascading effect within the service chain.

To address this, the Circuit Breaker Pattern, explained in \cite{cbp}, prevents the application from performing operations likely to fail. This safety mechanism can be crucial in maintaining system stability and preventing further complications. Both the Retry and Circuit Breaker patterns can be implemented directly within the application or through the infrastructure using service meshes like Istio and Traefik.

\paragraph{Column Schema Replication Pattern} This pattern involves replicating columns to make data available across different bounded contexts, as discussed in \cite{sathp}. On one hand, this strategy enhances performance and eliminates dependencies on other services when accessing data. On the other hand, maintaining data consistency and ensuring synchronization across services can pose significant challenges. Furthermore, implementing the replication mechanism — whether synchronous or asynchronous, requires careful consideration. The asynchronous approach is often preferable as it enhances system responsiveness and reduces availability issues by minimizing dependencies among services.

\paragraph{Replicated Caching Pattern} While caching generally improves performance by allowing data access from memory rather than disk, it can also facilitate access to distributed data. In this pattern, if service A needs to access data from service B, it retrieves this data directly from a cache that contains all data from service B, which are written to the cache upon changes and then replicated. This approach significantly enhances scalability and fault tolerance. However, it involves trade-offs, including service dependency. For instance, as the data owner, service B must populate the cache before service A can begin operations. Additionally, managing large volumes of data can be problematic; according to \cite{sathp}, 500 MB is considered a threshold. Architects must carefully consider the volume of data, cache size, and the number of service instances that require the cache. Furthermore, cache replication might experience delays in synchronizing data between services, making this pattern less suitable for data that changes frequently but more appropriate for data that changes less often. Finally, configuring and managing this pattern are complex tasks that require the services to be aware of each other.

\paragraph{Data Domain Pattern} Although sharing data is generally not recommended, this pattern offers benefits that may be worthwhile depending on the context, such as decoupled services which enhance availability as an architectural characteristic. Responsiveness is another significant advantage, as it eliminates the complications associated with the Interservice Communication Pattern; for example, the required data is just an SQL query away. This pattern also maintains data integrity and consistency, and allows for the use of RDBMS features such as views, procedures, and triggers.

Contrary to the Interservice Communication Pattern and Replicated Caching Pattern, the Data Domain Pattern does not involve contracts or abstract layers over the database, which makes changing database schemas more challenging as it requires coordination across all services within the bounded context. Additionally, there are heightened security and auditing risks, given that services have access to the entire database.

\subsection{Security and Auditing}
Security is a fundamental and non-negotiable architectural characteristic, especially given the millions of transactions that occur daily on the internet, many of which are fraudulent or non-secure. Both architectural styles must implement security measures, though the complexity and effort involved can vary. In monolithic architectures, there are relatively few entry points into the application, and once within the application layer, there are no additional security measures required among components to enable communication. However, \cite{msa} highlights that the attack surface is broader in microservice-based architectures, thereby increasing the risk. Distributed security can also impact performance due to the need to connect with a remote security token service, for instance.

Sharing user context is another concern; in a monolithic style, all components share the same session, whereas distributed architectures often use JWTs — a popular choice, to share user context among services, as described by \cite{msa}.

Furthermore, data security and auditing are critical. Encryption is required both in transit and at rest. Distributed architectures naturally have more components to authenticate, authorize, encrypt, and so on, increasing the security complexity.

\subsection{Coordinating distributed operations}
In contrast to monolithic, distributed-based styles requires a mechanism that enables each component to excecute its task. This mechanism can be either an orchestrator or a choreograpy among the services themselves.

\paragraph{Orchestration Style} An orchestrator is a component responsible for managing the workflow state, error handling, and notifications within a domain. \cite{sathp} makes an important point: microservices-based architectures typically use an orchestrator for each workflow because relying on a single orchestrator, such as in an Enterprise Service Bus, is discouraged due to the potential for undesirable coupling. The Orchestration Pattern offers excellent error handling, centralized workflow management, recovery capabilities, and state management. However, like any other pattern, it has its disadvantages. Responsiveness can be compromised since all requests must pass through the orchestrator, potentially creating a bottleneck—as was observed in the Amazon Prime case detailed in \cite{spvrc}. Moreover, fault tolerance is a concern because the orchestrator itself can become a single point of failure. Scalability can also be challenging due to the orchestrator's numerous coordination points, and the pattern inherently increases service coupling.

\begin{figure}[H]
\centering
\begin{tikzpicture}[scale=0.5, node distance=2.5cm, thick, align=center,
    orchestrator/.style={rectangle, draw, text width=6em, text centered, minimum height=3em},
    service/.style={rectangle, draw, text width=4em, text centered, minimum height=3em},
    line/.style={draw, -latex', thick},
    node distance=1cm]

    \node[orchestrator] (orchestrator) {Orchestrator};

    \node[service, below left=of orchestrator] (service1) {Service 1};
    \node[service, below=of orchestrator] (service2) {Service 2};
    \node[service, below right=of orchestrator] (service3) {Service 3};

    \draw[line] (orchestrator) -- (service1) node[midway, left] {};
    \draw[line] (orchestrator) -- (service2) node[midway, right] {};
    \draw[line] (orchestrator) -- (service3) node[midway, right] {};

\end{tikzpicture}
\vspace{10pt}
\caption{Orchestration style topology}
\label{fig:ost}
\end{figure}
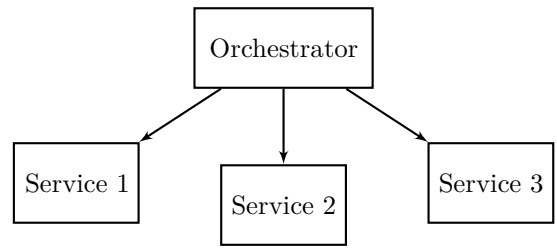

\paragraph{Choreography Style} In a choreography style, there is no central orchestrating component; instead, services communicate directly with each other. This approach enhances responsiveness and scalability by reducing the number of intermediaries in the workflow and potentially increasing parallelism. Fault tolerance is also improved due to the ability to scale services independently of orchestrator, which also reduces coupling by the same reason. However, distributing the workflow introduces challenges, particularly in error handling. Services must possess the knowledge typically centralized in an orchestrator, complicating both error resolution and efforts to enhance recoverability. Additionally, this style lacks centralized state management for ongoing operations, as delineated in \cite{sathp}.

\begin{figure}[H]
\centering
\begin{tikzpicture}[auto, node distance=4cm, thick, align=center,
    service/.style={circle, draw, text width=2.5em, text centered, minimum height=3em},
    arrow/.style={->, >=stealth, thick}]

    \node[service] (service1) {S1};
    \node[service, right of=service1] (service2) {S2};
    \node[service, below of=service1] (service3) {S3};
    \node[service, right of=service3] (service4) {S4};

    \draw[arrow] (service1) to[bend right] node[midway, left] {Event A} (service3);
    \draw[arrow] (service3) to[bend right] node[midway, right] {Event B} (service1);
    \draw[arrow] (service1) to node[midway, above] {Event C} (service2);
    \draw[arrow] (service3) to node[midway, below] {Event F} (service4);
    \draw[arrow] (service2) to node[midway, right] {Event D} (service4);
    \draw[arrow] (service4) to[bend right] node[midway, below] {Event E} (service3);

\end{tikzpicture}
\vspace{10pt}
\caption{Choreography style topology}
\label{fig:ost}
\end{figure}
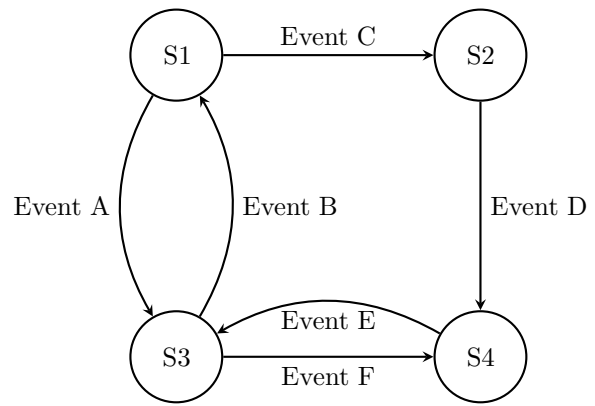

\subsection{Cost}
The widespread acceptance and popularization of distributed architectures were significantly propelled by the advent of cloud computing. The ease of deploying a virtual machine or container with just a few clicks — or even automatically through scriptable configurations, made it irresistible to explore the vast array of cloud products available. This includes everything from message brokers and various types of databases to load balancers, all accessible to businesses of any size. Thanks to the cloud's elasticity model, billing is based only on the resources actually used. However, costs are incurred in every aspect of usage, including data transfer, storage, access, requests, communication, and authorization, calculated down to the bit level.

The Amazon Prime case revealed by \cite{spvrc}, documents a dramatic 90\% cost reduction achieved by shifting back to a monolithic architecture. According to \cite{btfmm}, cost is the most common reason companies revert from microservices to monoliths. Operational expenses become unsustainable at a certain scale as each service instance may require its own server (or container), maintain a database, and generate separate metrics and logs.

\subsection{Observability}
Regardless of the architectural style employed, observability is critical for any serious enterprise application. Merely collecting logs, metrics, and traces is not sufficient, though these are foundational for testing and debugging systems. In \cite{dso}, observability is described as a property of a system that acknowledges the following realities:

\begin{enumerate}
    \item \textit{No complex system is ever fully healthy.}
    \item \textit{Distributed systems are pathologically unpredictable.}
    \item \textit{It’s impossible to predict the myriad states of partial failure various parts of the system might end up in.}
    \item \textit{Failure needs to be embraced at every phase, from system design to implementation, testing, deployment, and, finally, operation.}
    \item \textit{Ease of debugging is a cornerstone for the maintenance and evolution of robust systems.}
\end{enumerate}

In \cite{sre} the Four Golden Signals of monitoring are defined as latency, traffic, errors, and saturation. The complexity of monitoring increases linearly with the number of services and exponentially with the addition of components such as data synchronization tools. Moreover, if you implement patterns like Retry and Circuit Breaker within a Sidecar Pattern - which isolates components in containers, the complexity continues to rise.

As systems become more complex, observability becomes increasingly challenging. In distributed systems, tracing represents a major hurdle. It involves tracking a sequence of causally related events that describe the flow of a request through the system, represented as a directed acyclic graph of spans, as \cite{dso} delineates. While one benefit of a distributed architecture is the ability to support polyglot systems, which relies in the interoperability among diverse systems, this same diversity makes tracing more difficult. The challenge arises from the need to instrument a variety of frameworks, languages, and libraries.

\section{Conclusion}
There is always a trade-off involved with every decision. When an architect encounters a situation where no trade-off seems apparent, it likely has not yet been revealed. An architect's job entails remaining unbiased by the advantages of any particular architectural style or pattern and not aiming solely to enhance all characteristics. Instead, architects must be keenly aware of which disadvantages could be prohibitive due to business constraints. A clear evaluation process is essential, one that distinguishes what is critical, what is fundamental, and what is merely nice to have.

Recently, there has been a trend towards reverting to monolithic architectures due to concerns over costs, complexity, and performance, among other reasons. However, just as the previous enthusiasm for shifting towards microservices should have met with caution, the current trend of moving back to monoliths should also be approached carefully. There is no one-size-fits-all solution, as businesses vary and have their own particularities. Choosing an architectural style is not a question of right or wrong; styles can even be combined to achieve specific goals. It’s important to understand the business needs and consider the foreseeable future, while remaining grounded in reality.

Research methodologies that support decisions by evaluating business constraints in contrast to trade-offs among architectural styles are crucial. Additionally, investigating how to improve these architectural styles by combining the strengths of one with the weaknesses of another is fundamental to the evolution of software engineering.

\bibliographystyle{apalike-sol}
\bibliography{refs}

\end{document}